\documentclass[12pt]{iopart}

\usepackage{subcaption}
\usepackage{graphicx}

\begin{document}

\title[]{Highly reflective and high-$Q$ thin resonant subwavelength gratings}

\author{Gurpreet Singh$^1$, Trishala Mitra$^1$, S{\o}ren P. Madsen$^2$, and Aur\'e lien Dantan$^1$}

\address{$^1$Department of Physics and Astronomy, Aarhus University, DK-8000 Aarhus C, Denmark}
\address{$^2$Department of Mechanical and Production Engineering, Aarhus University, DK-8000 Aarhus C, Denmark}
\ead{dantan@phys.au.dk}
\vspace{12pt}
\begin{indented}
\item[]
\end{indented}

\begin{abstract}
We theoretically investigate the design of thin subwavelength gratings possessing high-reflectivity and high-$Q$ resonances when illuminated at normal incidence by a Gaussian beam. We compare the performances of single-period and dual-period rectangular gratings using Finite Element Method-based optimization and predict one to two orders of magnitude improvement in their transmission loss-linewidth product, which is the relevant figure of merit for e.g. resonant mirror-based microcavity applications.
\end{abstract}

%
%
%
%
%

\section{Introduction}

Resonant guided-mode gratings exploit the coupling between incoming out-of-plane radiation and transverse guided-modes in a periodic subwavelength structure in order to tailor the optical properties of thin films. This makes them ubiquitous in a wide range of photonics, sensing, lasing and optomechanics applications~\cite{Wang1993,Rosenblatt1997,ChangHasnain2012,Quaranta2018}. Many of these applications benefit from the achievement of high-quality factor optical Fano resonances in the reflection and/or transmission spectrum of these gratings~\cite{Limonov2017}. In applications in which the mechanical vibrations of thin suspended films are exploited for optical sensing and optomechanics purposes, it is generally desirable to operate with ultrathin (few tens to a few hundreds of nanometers), suspended films made of low-loss material, e.g. silicon nitride. In this case patterning these thin films with one-dimensional subwavelength gratings or two-dimensional photonic crystal structures allows for strongly enhancing their otherwise low reflectivity while preserving their low mass and high mechanical quality~\cite{Kemiktarak2012,Bui2012,Norte2016,Reinhardt2016,Chen2017,Nair2019,Darki2022}. Typically, though, the quality factor $Q$ of the achieved optical resonances is moderate, either by design (broadband reflectivity applications) or because of the inherently low modulation index. 

However, achieving high-reflectivity and high Q-factor resonances can be very beneficial for applications involving the use of these mirrors in optomechanical cavities~\cite{Xuereb2012,Kemiktarak2012b,Chen2017,Naesby2018,Gartner2018,Cernotik2019,Manjeshwar2020,Fitzgerald2021,Xu2022,Sang2022,Zhou2023,Enzian2023,Manjeshwar2023}. Indeed, it was recently proposed~\cite{Naesby2018} and demonstrated~\cite{Mitra2024} that ultrashort cavities consisting in the plane-plane arrangement of such a suspended resonant mirror and a conventional broadband mirror may exhibit substantially lower linewidths/higher $Q$s than a conventional broadband mirror cavity with the same length and internal losses. Such optomechanical "Fano" microcavities are then attractive for investigating novel optomechanical phenomena~\cite{Cernotik2019,Fitzgerald2021,Manjeshwar2023,Peralle2023}, realizing optomechanical bistability-based photonic devices~\cite{Sang2022,Zhou2023,Enzian2023} or optomechanical sensors~\cite{Naserbakht2019,AlSumaidae2021,Horning2022,Salimi2023}.

In this work, motivated among others by recent realizations and applications using thin suspended silicon nitride films~\cite{Nair2019,Parthenopoulos2021,Darki2021,ToftVandborg2021,Darki2022,Darki2022b,Mitra2024}, we theoretically investigate the design of thin resonant subwavelength gratings possessing high-reflectivity and high-$Q$ resonances when illuminated at normal incidence by a Gaussian beam, as depicted in Fig.~\ref{fig:schematic}(a). For such thin subwavelength gratings operating in the ''weak-contrast" regime, the finite size of the grating/beam as well as beam collimation effects are known to strongly affect the guided-mode resonances, intrinsically limiting the achievable resonance linewidths and peak reflectivity levels~\cite{ToftVandborg2021}. Indeed, for a canonical grating such as the one shown in Fig.~\ref{fig:schematic}(b), consisting of a single rectangular grating finger per unit cell, it is well-known that a narrow spectral linewidth is intrinsically associated with a small angular tolerance~\cite{Wang1993,Rosenblatt1997}. A narrow-linewidth grating with a weak index modulation will thus require an area and an illuminating beam size larger than the propagation length of the guided wave in the grating in order to have an efficient interference between the incoming light field and the guided-mode and, thereby, avoid linewidth broadening and reduced peak reflectivity~\cite{Boye2000,Jacob2001,Bendickson2001,ToftVandborg2021}. As proposed in~\cite{Lemarchand1998,Mizutani2003} and demonstrated in~\cite{Fehremback2007}, it is nevertheless possible to realize spectrally \textit{narrow} resonant grating filters with a \textit{broad} angular tolerance by considering doubly-periodic structures, e.g. such as the one depicted in Fig.~\ref{fig:schematic}(c) whose unit cell consists of two rectangular fingers with slightly different finger widths.

We thus investigate the occurence of high-reflectivity, high-$Q$ resonances in such dual-period gratings, compare their performances with corresponding single-period gratings and perform an optimization of their design based on Finite Element Method (FEM) simulations. While the optimization strategy is generally applicable to a wide range of structures and applications we illustrate it with typical parameters inspired from currently investigated Fano microcavities~\cite{Naesby2018,Cernotik2019,Manjeshwar2020,Manjeshwar2023,Mitra2024} and predict a one to two orders of magnitude improvement in the relevant figure of merit (transmission loss-linewidth product).

The paper is organized as follows: in Sec.~\ref{sec:singlevsdual} a comparison of the resonances of a single-period and a dual-period grating is given as an example, first on the ideal basis of infinite structures illuminated by plane waves, then on the realistic basis of FEM simulations of finite-size gratings under Gaussian beam illumination. In Sec.~\ref{sec:optimization} a systematic investigation based on FEM simulations is performed as a function of the grating vertical and horizontal fill ratios in order to identify the optimal parameters for given grating type and beam size. In Sec.~\ref{sec:algorithm} a fast optimization algorithm is proposed and implemented to investigate the effect of the grating finger width asymmetry on the achievable figures of merit, before we conclude in Sec.~\ref{sec:conclusion}.

\section{Single- {\it vs} dual-period grating}
\label{sec:singlevsdual}

\subsection{Principle}

\begin{figure}
\centering
\begin{subfigure}{0.49\columnwidth}
\includegraphics[width=\columnwidth]{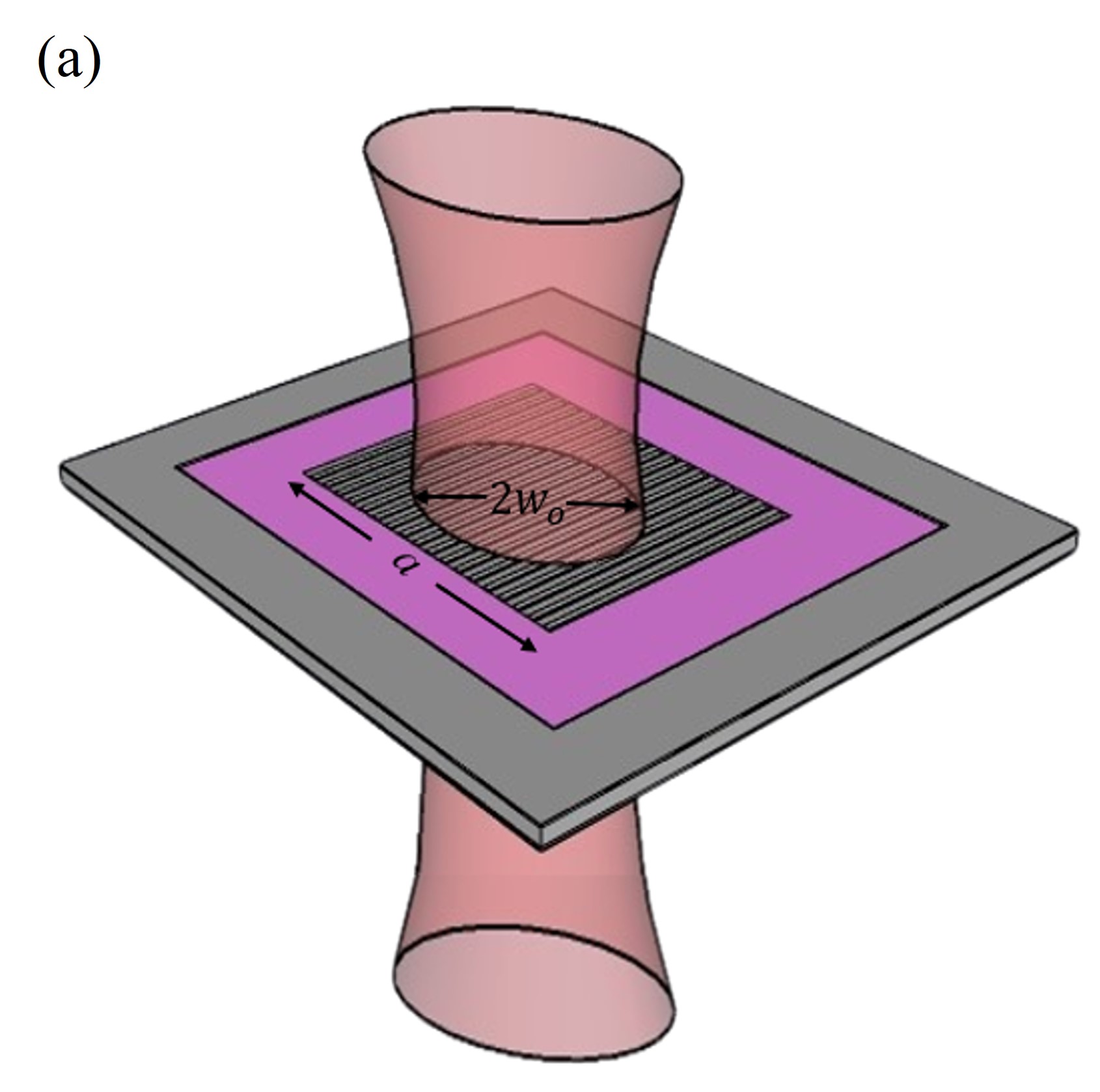}
\end{subfigure}
\begin{subfigure}{0.49\columnwidth}
\includegraphics[width=\columnwidth]{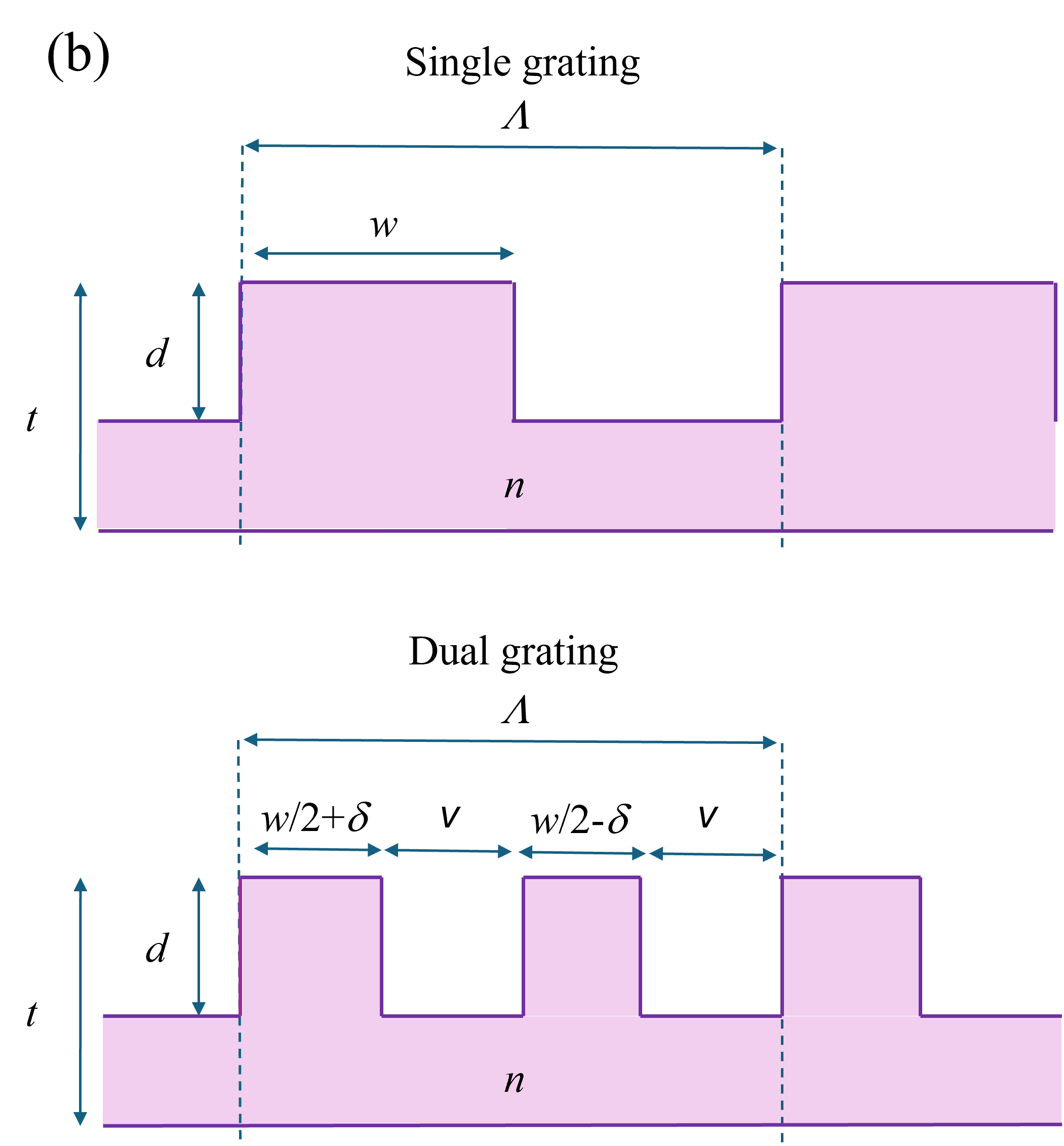}
\end{subfigure}
\caption{(a) Schematic illustration of the situation considered: a Gaussian beam with waist $w_0$ impinges at normal incidence on a subwavelength grating with transverse size $a$. (b) Transverse cross-sections of the single- and dual-period gratings considered.}
\label{fig:schematic}
\end{figure}

We consider the situation depicted in Fig.~\ref{fig:schematic}: a subwavelength grating is illuminated by a Gaussian beam with waist (radius) $w_0$, at or close to normal incidence. The incoming light is assumed to be monochromatic with wavelength $\lambda$ and TE-linearly polarized. The two grating configurations considered are represented in Fig.~\ref{fig:schematic}; both gratings consist in rectangular fingers with height $d$ supported by a substrate of the same lossless dielectric material with thickness $d'$. We denote by $t=d+d'$ the total film thickness and $n$ the refractive index of the material, assumed real and independent of the wavelength in the range considered. The grating is assumed to be thin with respect to the wavelength, so one operates in the low-constrast regime. The \textit{single-} and \textit{dual-}period gratings of Fig.~\ref{fig:schematic} differ by their unit cell (length $\Lambda$), which consists of a single finger with width $w$ in the case of the single-period grating, and  of two fingers with widths $w/2+\delta$ and $w/2-\delta$ separated by the same gap $v=(\Lambda-w)/2$ in the case of the dual-period grating.

\subsection{Plane-wave illumination}
\label{sec:planewave}

\begin{figure}
\centering
\begin{subfigure}{0.49\columnwidth}
\includegraphics[width=\columnwidth]{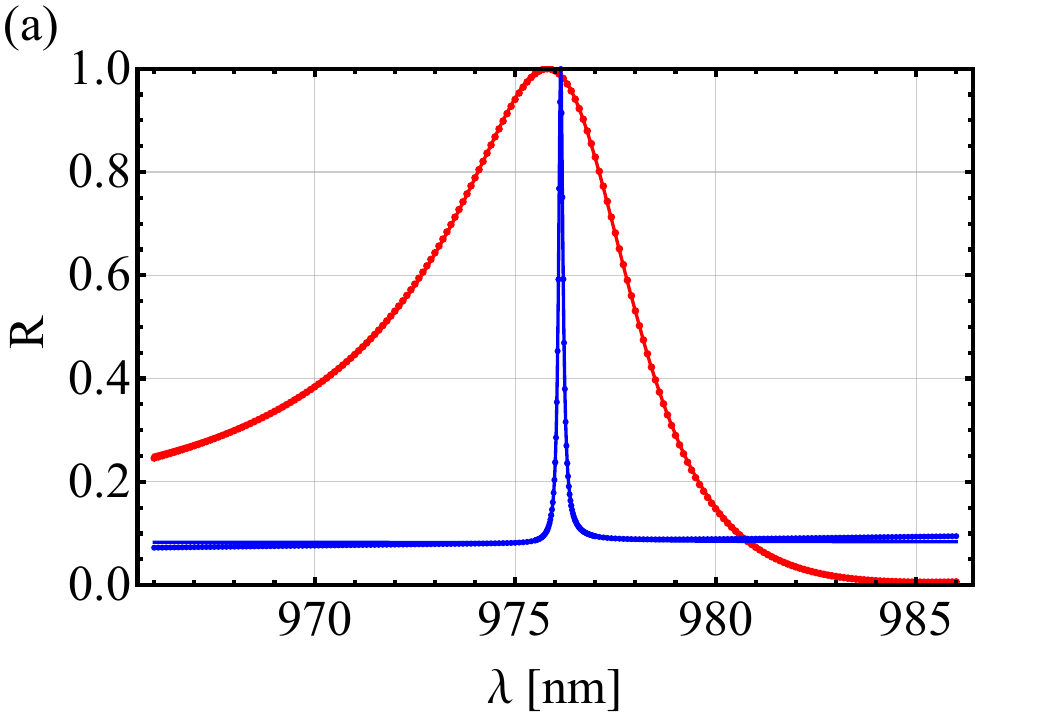}
\end{subfigure}
\begin{subfigure}{0.49\columnwidth}
\includegraphics[width=\columnwidth]{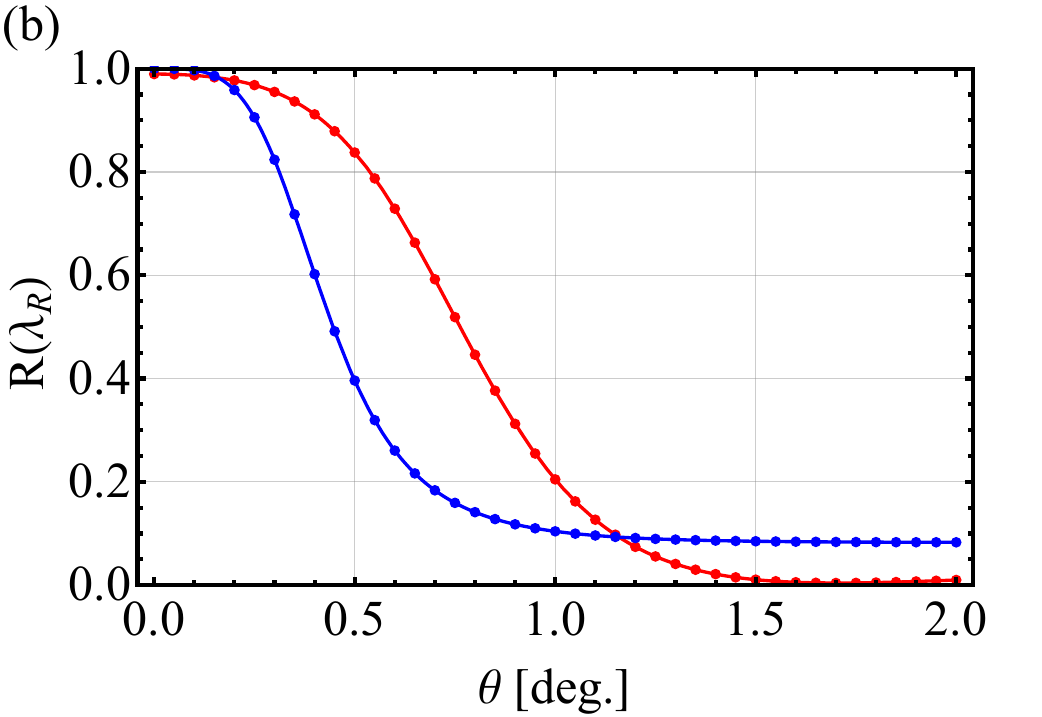}
\end{subfigure}
\caption{(a) Normal incidence reflectivity of the single-period (red points) and dual-period (blue points) gratings described in the text (Tab.~\ref{tab:mist_parameters}). (b) Peak reflectivity, evaluated at the normal incidence resonant wavelength $\lambda_R$, as a function of the incidence angle $\theta$ for both gratings. The solid lines in (a) are the result of fits to the coupled-model predictions (Eq.~(\ref{eq:Rmist})). In (b) they show the results of interpolations, both to guide the eye and assess the angular linewidth.}
\label{fig:mist}
\end{figure}

To illustrate the fact that much narrower Fano resonances can be obtained with a dual-period grating than with a single-period grating possessing a similar horizontal fill ratio $w/\Lambda$, Fig.~\ref{fig:mist} shows the results of Rigorous Coupled Wave Analysis simulations of their reflectivity assuming an infinite structure illuminated by a TE-polarized plane wave~\cite{MIST}. The total film thickness and refractive index are assumed to be $t=200$ nm and $n=2.3$ in both cases. The grating parameters reported in Table~\ref{tab:mist_parameters}, close to typical parameters used in recent investigations~\cite{Nair2019,Parthenopoulos2021,Darki2021,ToftVandborg2021,Darki2022,Darki2022b,Mitra2024}, were chosen so as to achieve a guided-mode resonance at approximately the same wavelength ($\sim976$ nm), as well as to being close to the optimal parameters in terms of minimizing the Fano resonance linewidth and maximizing the peak reflectivity, as will be discussed in the next sections. The normal incidence spectra of Fig.~\ref{fig:mist}(a) are well-reproduced by a "Fano"-resonant reflectivity expression based on a simple couple-mode model~\cite{Bykov2015,Parthenopoulos2021}
\begin{equation}
R=\left|r_d\frac{k-k_0}{k-k_1+i\gamma}\right|^2,
\label{eq:Rmist}
\end{equation}
where $r_d$ is the off-resonant reflectivity coefficient, $k=2\pi/\lambda$ the incident light wavenumber, $k_0=2\pi/\lambda_0$ and $k_1=2\pi/\lambda_1$ are the zero-reflectivity wavenumber and the Fano resonance pole wavenumber, respectively, and $\gamma$ is related to the spectral linewidth (FWHM) in wavelength $\Delta\lambda$ of the Fano resonance by $\Delta\lambda=(\lambda_0^2/\pi)\gamma$. Figure~\ref{fig:mist}(b) shows the variation of the peak reflectivity for small oblique incidence angles, which allows to extract the angular linewidth $\Delta\theta$. The spectral and angular linewidths for both gratings are reported in Tab.~\ref{tab:mist_parameters} and clearly illustrate that a much narrower spectral linewidth is observed with the dual-period grating while both gratings exhibit similar angular linewidths. One may then expect a similar sensitivity to finite-size and collimation effects, and thereby a similar peak reflectivity for both gratings, in the realistic scenario of a finite-size grating illuminated by a focused Gaussian beam, which we now turn to.

\begin{table}
\caption{Single- and dual-period grating parameters.}
\centering
  \label{tab:mist_parameters}
  \begin{tabular}{c|ccccc|cc}
   Grating & $\Lambda$ (nm) &  $t$ (nm) & $d$ (nm) & $w$ (nm) & $\delta$ (nm) &  $\Delta\lambda$ (nm) & $\Delta\theta$ (deg.)\\\hline
   Single & 539 & 200 & 40 & 376 & - & 5.46 & 1.52\\
   Dual & 532 & 200 & 150 & 370 & 10 & 0.13 & 0.90\\
  \end{tabular}
\end{table}

\subsection{Finite-size and collimation effects}

\begin{figure}
\centering
\begin{subfigure}{0.49\columnwidth}
\includegraphics[width=\columnwidth]{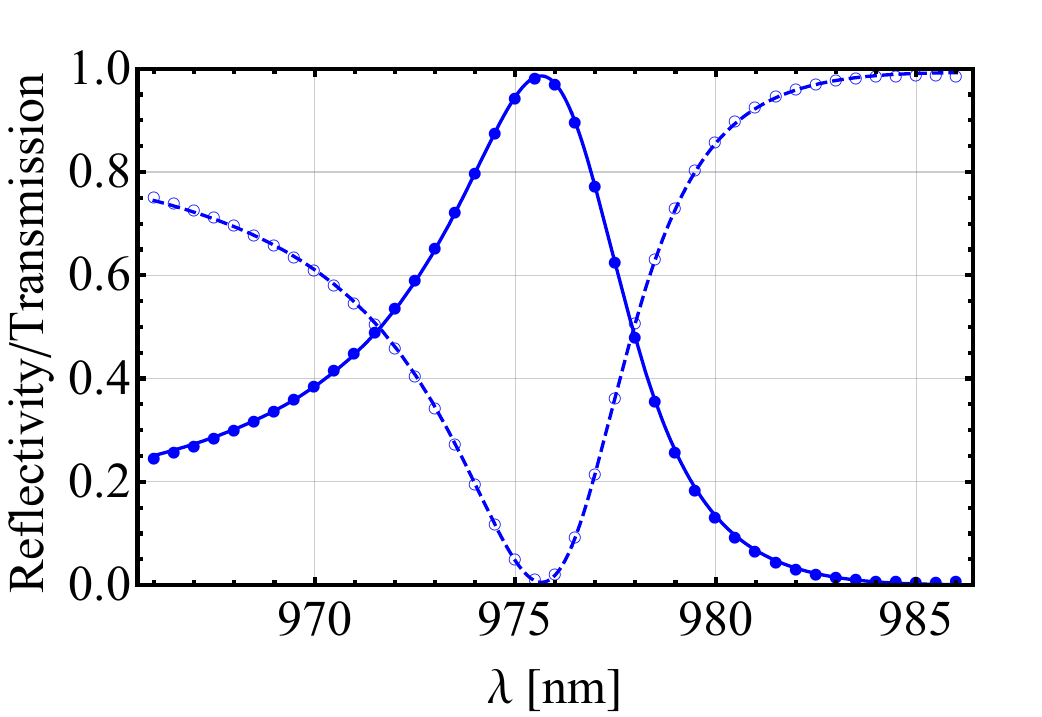}
\end{subfigure}
\begin{subfigure}{0.49\columnwidth}
\includegraphics[width=\columnwidth]{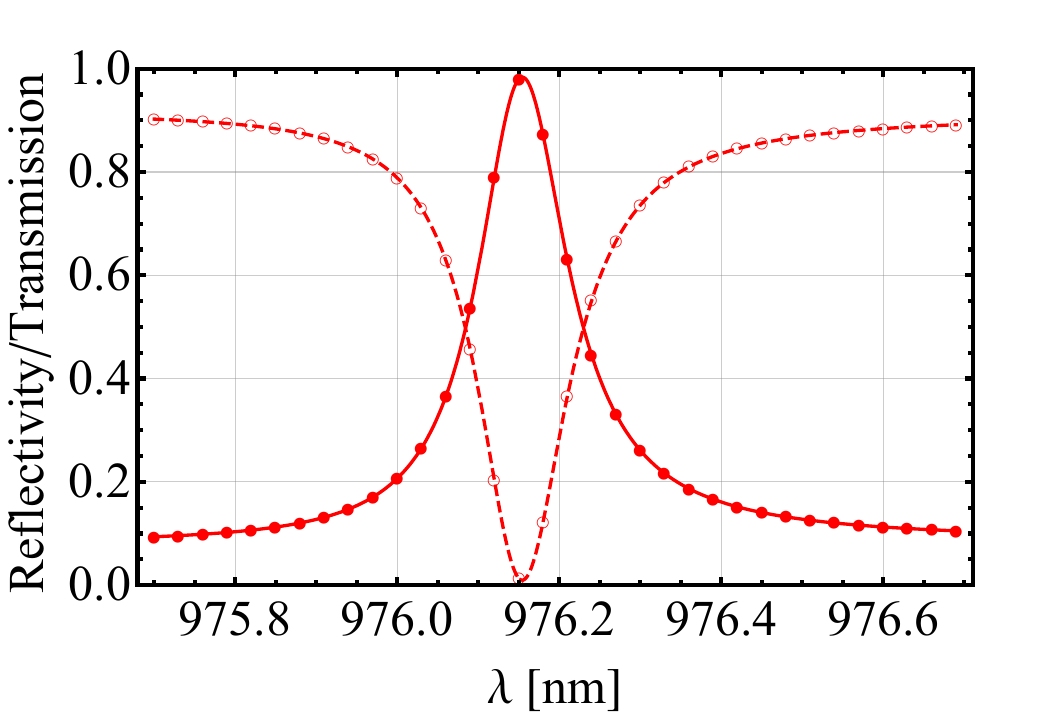}
\end{subfigure}

\begin{subfigure}{0.49\columnwidth}
\centering\includegraphics[width=0.8\columnwidth]{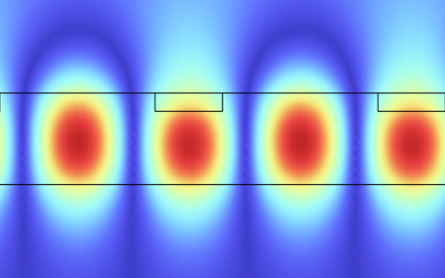}
\end{subfigure}
\begin{subfigure}{0.49\columnwidth}
\centering\includegraphics[width=0.8\columnwidth]{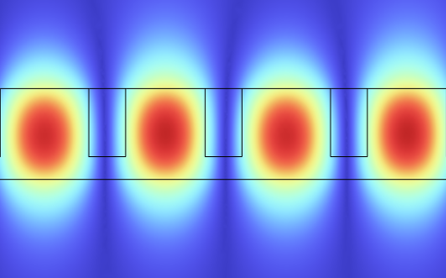}
\end{subfigure}
\caption{Reflectivity (filled dots) and transmission (empty dots) spectra of the single-period (top left) and dual-period (top right) gratings of Fig.~\ref{fig:mist} for a Gaussian beam waist of 50 $\mu$m. The plain and dashed lines are the results of fits to the coupled-mode model including finite-size and collimation effects. The bottom left and right figures show the electric field intensity distribution at resonance (976 nm) for the single- and dual-period grating, respectively.}
\label{fig:comsol_spectra}
\end{figure}

To model the realistic situation of a finite-size grating illuminated by a focused Gaussian beam we performed Finite Element Method simulations of a two-dimensional grating (infinite in the grating finger direction) using the COMSOL Multiphysics environment. In order to model a situation close to current experiments while keeping the simulations tractable, a transverse grating size of $a=135$ $\mu$m and waists of a few tens of microns were assumed. Figure~\ref{fig:comsol_spectra} shows the transmission and reflectivity spectra of the single- and dual-period gratings of the previous section for an incoming TE-polarized beam with a waist of 50 $\mu$m. Following the coupled-mode model of Ref.~\cite{ToftVandborg2021}, which was shown in  to reproduce well the experimentally measured spectra of this type of gratings in presence of finite-size and collimation effects, the reflection spectra of Eq.~(\ref{eq:Rmist}) can be modified by substituting $k_0$ with $k_0+i\beta$, where $\beta$ is a real constant accounting for the losses due to finite-size and collimation effects. Fits to the model give linewidths of 5.46 nm and 0.13 nm for the single- and dual-period gratings, respectively, showing negligible broadening as compared to the infinite structure/plane wave situation investigated in Sec.~\ref{sec:planewave}. However, non-unity peak reflectivity of 98.6\% and 98.3\%, respectively, are now observed as a result of collimation and finite-size effects, as expected. As abovementioned, such a non-unity peak reflectivity is particularly relevant for the realization of high-finesse Fano microcavities exploiting high-$Q$ resonant mirrors. As predicted in~\cite{Naesby2018} and experimentally observed in~\cite{Mitra2024}, the spectral linewidth of such Fano cavities is indeed ultimately limited by the product of the cavity losses and the resonant mirror linewidth. In this context, a natural figure of merit for optimizing the design of the resonant mirror is the product of its $Q$-factor, $\lambda_R/\Delta\lambda$, divided by the resonant mirror transmission losses at resonance, $1-R_\textrm{max}$, 
\begin{equation}
\textrm{FOM}=\frac{Q}{1-R_\textrm{max}}, \label{eq:FOM}
\end{equation}
as this quantity will represent, up to a constant of order unity, the $Q$-factor of the Fano cavity consisting of the Fano mirror and a perfectly reflecting broadband mirror~\cite{Mitra2024}.

\begin{figure}
\centering
\includegraphics[width=\columnwidth]{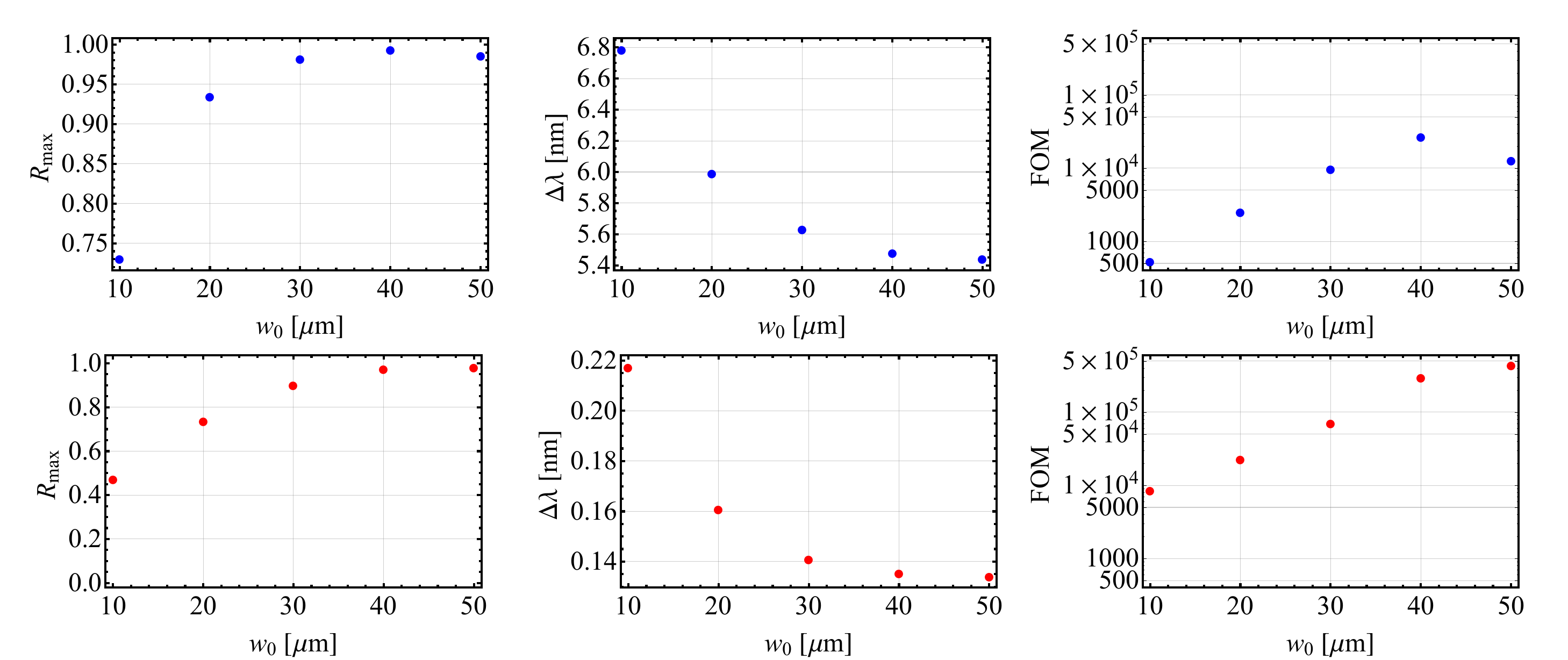}
\caption{Variation with the waist $w_0$ of the incident beam of the peak reflectivity $R_\textrm{max}$, spectral linewidth $\Delta\lambda$ and the figure of merit of Eq.~(\ref{eq:FOM}) for the single-period grating (top row) and dual-period grating (bottom row) of Fig.~\ref{fig:comsol_spectra}.}
\label{fig:comsol}
\end{figure}

Figure~\ref{fig:comsol} shows the variations of the peak reflectivity $R_\textrm{max}$, the spectral linewidth $\Delta\lambda$ and the previously introduced figure of merit of both gratings as a function of the Gaussian beam waist. For large waists, one sees that the dual-period grating linewidth is more than 40 times narrower than that of the single-period grating, while their peak reflectivities are similar. This means that, even though the coupling with the guided mode is weaker, a stronger field localization is achieved in the dual-period grating, leading to the same reflectivity loss level. It can also be seen that, as the beam waist is decreased, the linewidth broadening is comparable and relatively moderate for both gratings. In contrast, the peak reflectivity strongly decreases as the waist is reduced. Accordingly, this leads to a degradation of the FOM by several orders of magnitude in the range of waists considered. Regardless of the waist chosen, though, the use of the double-period grating substantially improves the figure of merit as compared to the single-period grating.

\section{Systematic optimization using FEM simulations}
\label{sec:optimization}

\subsection{Single-period grating {\it vs} dual-period grating with $\delta=10$ nm}

\begin{figure}
\centering
\includegraphics[width=\columnwidth]{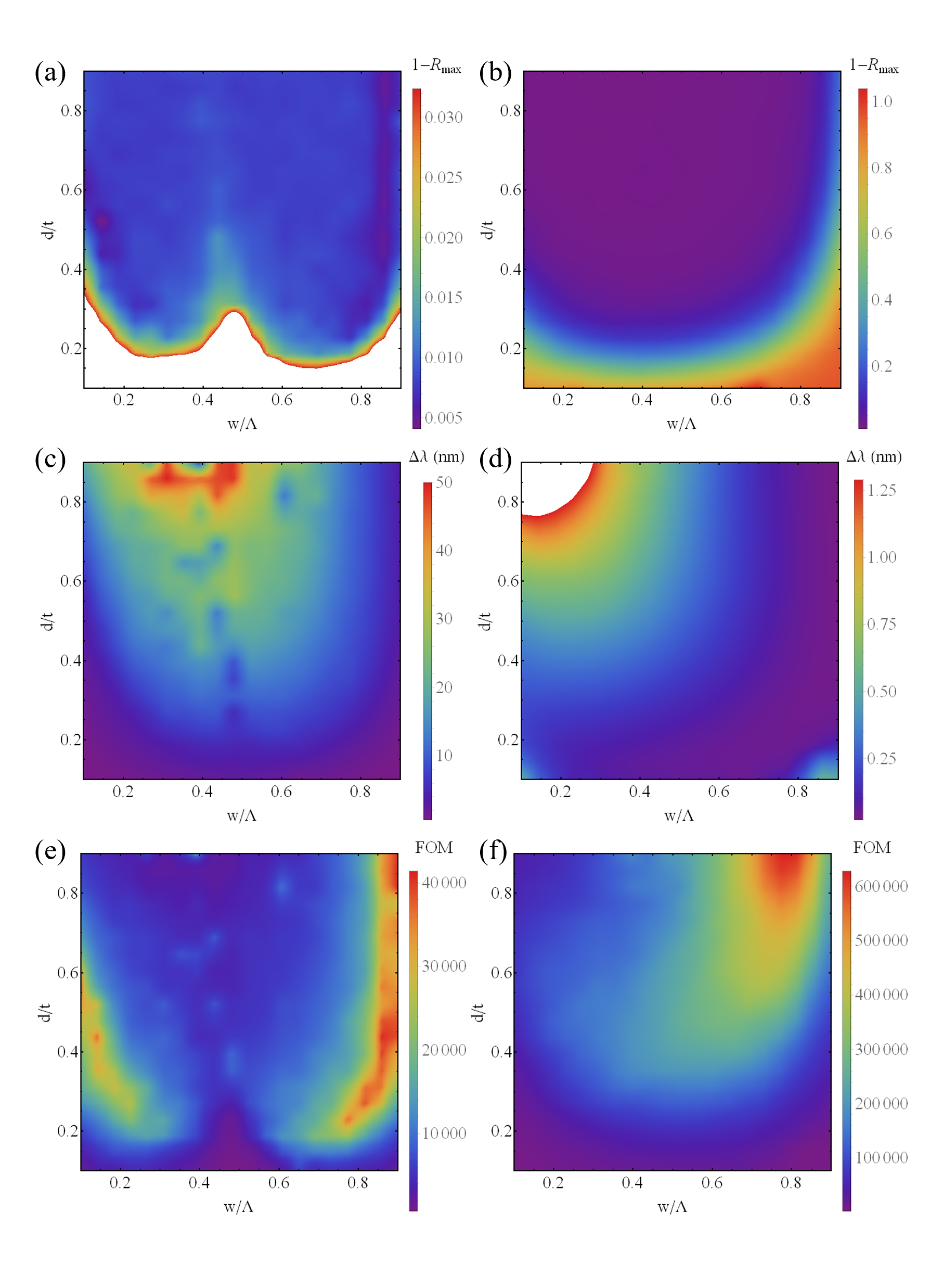}
\caption{FEM-simulated loss level $(1-R_\textrm{max})$, linewidth $\Delta\lambda$ and FOM for single-period (left column) and dual-period (right column) gratings as a function of their horizontal and vertical fill ratios, $w/\Lambda$ and $d/t$. Other parameters: $t=200$ nm, $n=2.3$, $\delta=10$ nm, $w_0=50$ $\mu$m. $\Lambda$ is adjusted to achieve a resonant wavelength of 976 nm.}
\label{fig:FEM_optimization}
\end{figure}

To perform a systematic comparison and optimize the design of the two types of gratings, FEM simulations were carried out with the following (experimentally motivated) constraints: first, the refractive index and total film thickness as well as the transverse size of the grating were fixed to the previous values. The incoming Gaussian beam waist was set to 50 $\mu$m and assumed to be TE-polarized. The horizontal and vertical grating fill ratios, $w/\Lambda$ and $d/t$, were varied in the range 10\%-90\% to avoid extreme values which could be challenging fabrication-wise. The grating period $\Lambda$ was adjusted so as to ensure a resonant wavelength close to 976 nm for all gratings. In a first series of simulations a conservative and fabrication-wise reasonable asymmetry parameter value of $\delta=10$ nm was assumed for the dual-period grating. 

The results of this systematic parameter space exploration are shown in Fig.~\ref{fig:FEM_optimization}. Figure.~\ref{fig:FEM_optimization}(a) shows that a peak reflectivity higher than 95\% can be obtained with a single-period grating for a wide range of the parameter space, in particular for deep grating fingers (high $d/t$ values). Figure~\ref{fig:FEM_optimization}(c) shows that low linewidths are in contrast achieved for shallow grating fingers and for either high or low horizontal fill ratios. This leads to maximal values of the FOM in the few $10^4$ range for shallow grating fingers with either high or low horizontal fill ratios (Fig.~\ref{fig:FEM_optimization}(e)). For instance, the parameters $\Lambda=539.4$ nm, $w=415.7$ nm, $d=45.9$ nm yield a peak reflectivity of 99.3\%, a linewidth of 3.9 nm and a FOM of $3.6\times 10^4$.

For the dual-period grating the peak reflectivity is also maximized for deep grating fingers (Fig.~\ref{fig:FEM_optimization}(b)), while the linewidth is observed to be minimized for high horizontal fill ratios (Fig.~\ref{fig:FEM_optimization}(d)). This leads to optimal FOM values for dual-period gratings with thick and deep grating fingers. For instance, $\Lambda=527.1$ nm, $w=406.3$ nm and $d=154.1$ nm yield $R_\textrm{max}\simeq 97.4$\%, $\Delta\lambda\simeq $ 0.076 nm and a FOM of $4.9\times 10^5$. In general, the dual-period grating's substantially narrower linewidths, together with peak reflectivities slightly lower but comparable to those of similar single-period gratings, allow for significantly improved figures of merit for the same grating/beam size.

\subsection{Variation with asymmetry parameter $\delta$}

\begin{figure}
\centering
\includegraphics[width=\columnwidth]{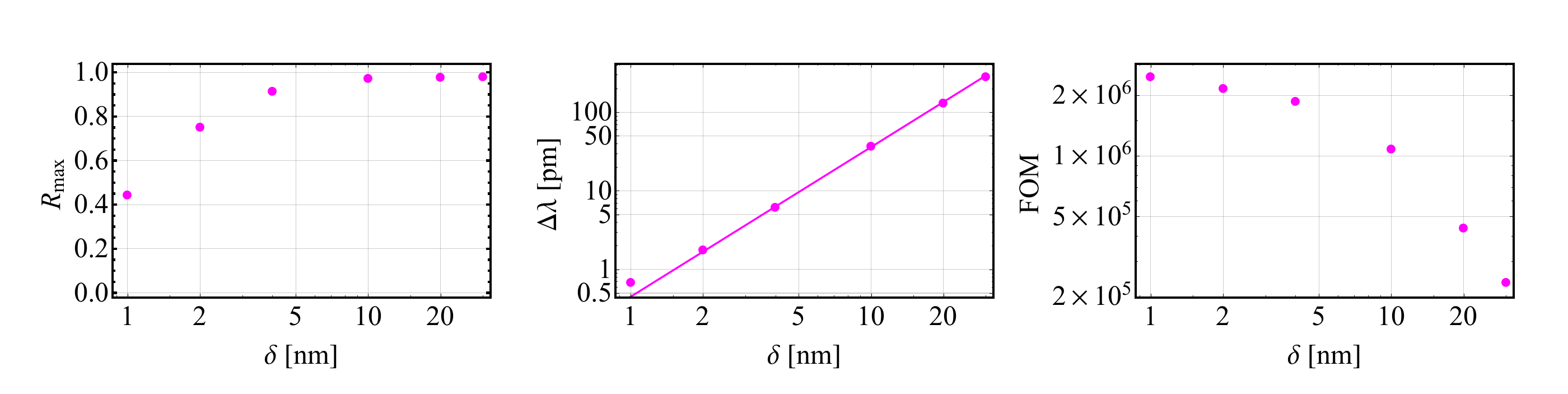}
\caption{FEM-simulated loss level $(1-R_\textrm{max})$, linewidth $\Delta\lambda$ and FOM for a dual-period grating with $\Lambda\simeq533$ nm, $w\simeq362$ nm and $d=150$ nm as a function of asymmetry parameter $\delta$. The line in the middle figure shows the result of a polynomial fit $\Delta\lambda=a\delta^\alpha$, yielding $\alpha=1.9$.}
\label{fig:delta}
\end{figure}

To illustrate the effect of the dual-period grating asymmetry parameter, Fig.~\ref{fig:delta} shows the variation of the peak reflectivity, the resonance linewidth and the FOM for a dual-period grating with $\Lambda\simeq 533$ nm, $w\simeq362$ nm, $d=150$ nm and varying $\delta$ (for each value of $\delta$, $\Lambda$ and $w$ were slightly adjusted so as to get a 976 nm resonant wavelength). It can be seen that extremely narrow linewidths can be obtained when the asymmetry parameter goes to 0, as the resonance becomes a high-$Q$ quasi-guided mode, which displays features similar to that of a bound state in the continuum~\cite{Sun2024}. In agreement with the findings of~\cite{Koshelev2018,Sun2024}, one observes in particular that the resonance linewidth scales approximately quadratically with the degree of asymmetry. However, this ultimately occurs at the expense of a decrease in the peak reflectivity, which may become too low for some cavity applications. This could be counterbalanced by an increase in the grating/beam size, which would reduce collimation and finite-size effects. In practice, though, fabrication inhomogeneities/size will ultimately limit the degree of asymmetry and thereby the achievable linewidth and FOM.

\section{Optimization algorithm}
\label{sec:algorithm}

Rather than scanning the whole parameter space as in Fig.~\ref{fig:FEM_optimization}, one can use an optimization algorithm to arrive at the parameters maximizing the FOM in Eq. (\ref{eq:FOM}). The evaluation of the FOM function involves the calculation of $R$ as a function of the wavelength to determine $1-R_\textrm{max}$ directly and $\Delta\lambda$ by curve fitting. Both quantities can contain numerical noise, resulting in small-scale oscillations of the FOM function. This suggests using a global optimiztion algorithm, as a local optimizer will quickly get stuck in a local minimum. We use here Bayesian optimization, implemented in the 'bayesopt' function in the statistics and machine learning toolbox in MATLAB R2023a \cite{MATLAB}. The algorithm works by initially sampling the FOM function at randomly chosen points and using the data to build a Gaussian process regression model. The model is then iteratively improved by adding more sampling points, biased towards the model optimum, until a maximum number of sample points has been reached. The optimum is finally taken as the best of the model optimum and the points visited in the iterative process. The final point determined by this method is not garanteed to be the global optimum, but generally shows very good FOM values. As the algorithm is stochastic, it is run several times.

\subsection{Convergence}
The algorithm can be used to directly determine the three parameters $\Lambda$, $w/\Lambda$ and $d/t$. It was found, however, that only optimizing with respect to the latter two parameters and determining the period by a 1D maximization of the reflectivity generally results in less computational time spent to locate a near-optimal solution, as compared to optimizing directly with respect to all three parameters.

\begin{figure}
\centering
\includegraphics[width=0.49\columnwidth]{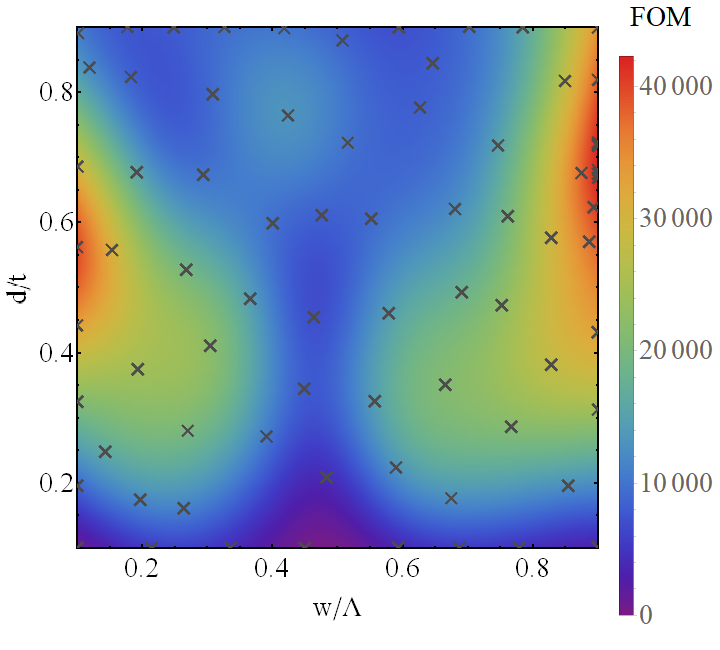}
\includegraphics[width=0.49\columnwidth]{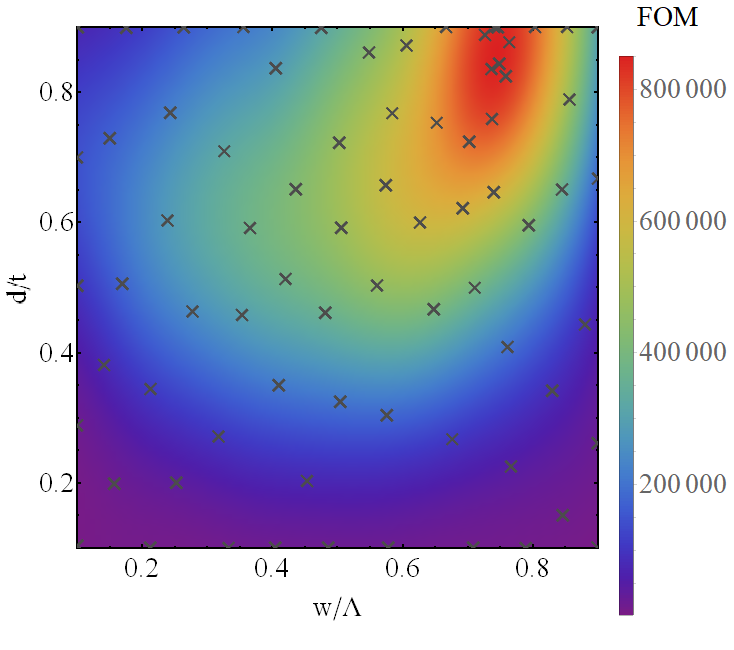}
\caption{The Gaussian process regression model resulting from running the optimization algorithm for the single-period (left) and dual-period (right) gratings. The crosses show the location of the sample points used. The best FOMs found by the algorithm are seen to be similar to the optimum ones found by the full parameter space scan of Fig.~\ref{fig:FEM_optimization}.}
\label{fig:model}
\end{figure}

The algorithm was tested against the data in Fig.~\ref{fig:FEM_optimization} and a maximum of 75 Bayesian optimization iterations was used with 20 of them being the initial non-biased random sampling. For both the single and dual period gratings, the algorithm was run a few times. In all cases, the Gaussian process regression models, examples of which are shown in Fig.~\ref{fig:model}, look similar to the data plotted in Fig.~\ref{fig:FEM_optimization}.

\subsection{Variation with $\delta$}

\begin{figure}
\centering
\begin{subfigure}{0.49\columnwidth}
\includegraphics[width=\columnwidth]{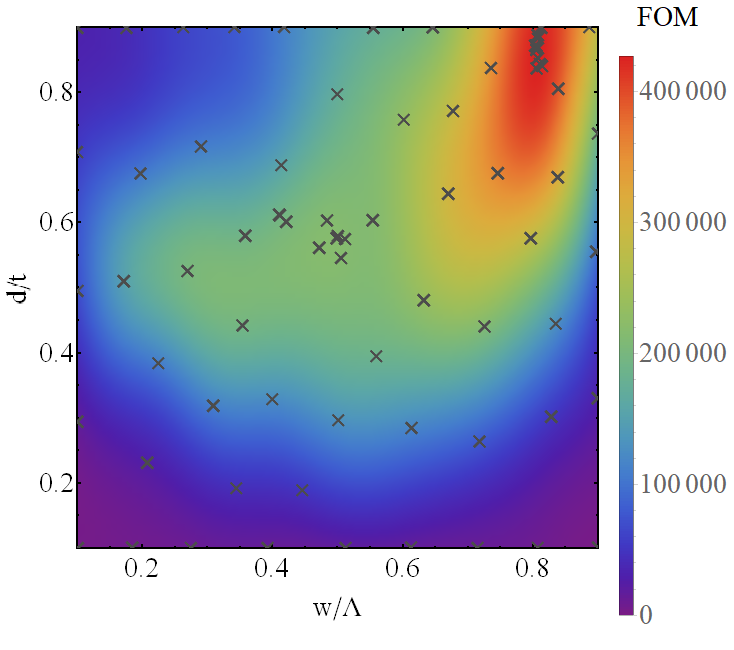}
\end{subfigure}
\begin{subfigure}{0.49\columnwidth}
\centering\includegraphics[width=\columnwidth]{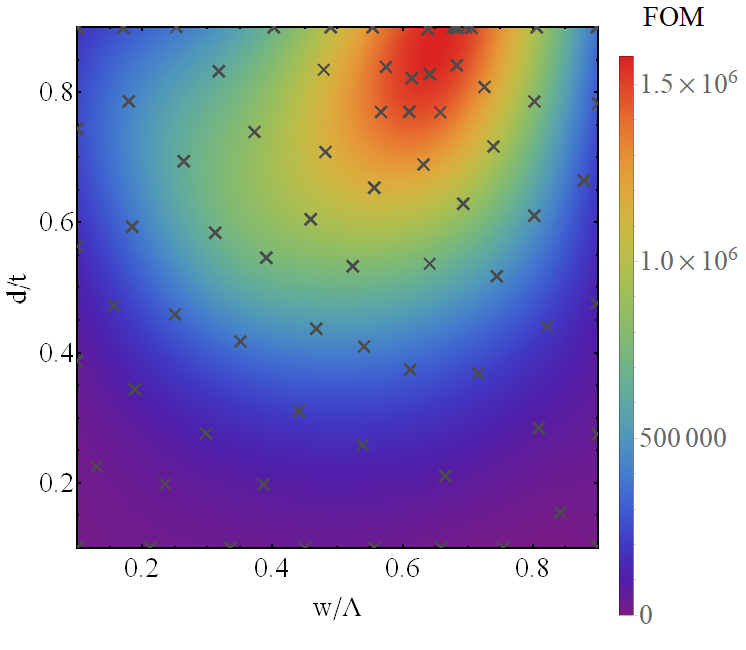}
\end{subfigure}
\begin{subfigure}{0.49\columnwidth}
\centering\includegraphics[width=\columnwidth]{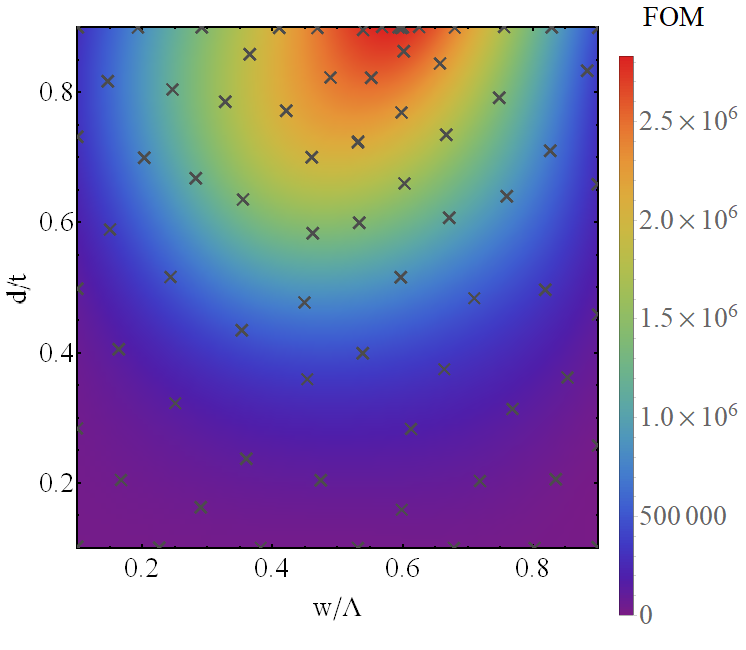}
\end{subfigure}
\begin{subfigure}{0.49\columnwidth}
\centering\includegraphics[width=\columnwidth]{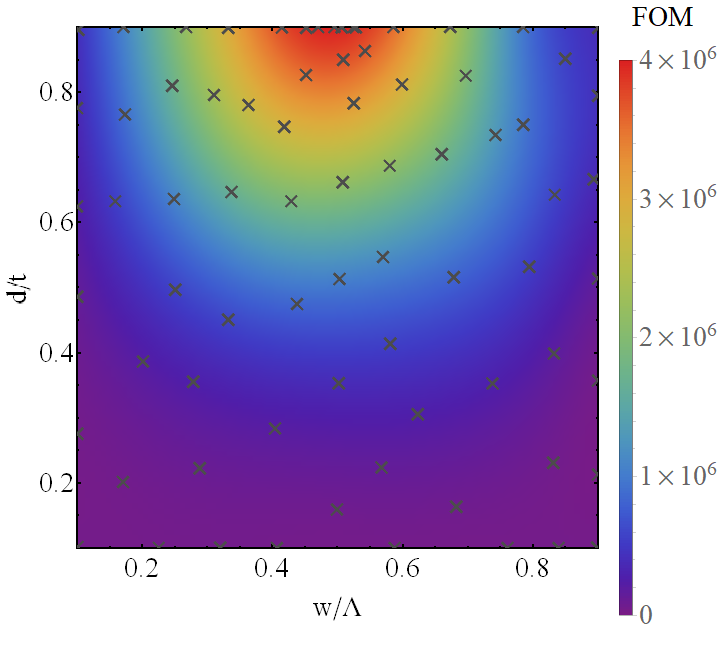}
\end{subfigure}
\caption{Gaussian process regression models resulting from running the optimization algorithm of the FOM for dual-period gratings with $\delta=20$ nm (top left), $\delta=5$ nm (top right), $\delta=2.5$ nm (bottom left) and $\delta=1.25$ nm (bottom right). The crosses show the location of the sample points used.}
\label{fig:opttot}
\end{figure}

The algorithm was used to find the near-optimal parameters for different values of the asymmetry parameter $\delta$. The optimization results are summarized in Tab.~\ref{tab:delta_study} and the resulting models are shown in Fig.~\ref{fig:opttot}. As observed in the specific example of Fig.~\ref{fig:delta}, the FOM value increases for decreasing $\delta$. As $\delta$ is decreased, the optimal horizontal fill ratio $w/\Lambda$ is also seen to move from $\sim 0.8$ to $\sim 0.5$, while the optimal vertical fill ratio $d/t$ stays close to 0.9, the maximum allowed value.

\begin{table}
\caption{Parameters found from running the optimization algorithm once for the single-period grating (first row) and for dual-period gratings with different $\delta$ values (other rows).}
\centering
\label{tab:delta_study}
\begin{tabular}{c|ccc}
$\delta$ (nm) & $d/t$ &  $w/\Lambda$ & FOM\\
\hline
- & 0.900 & 0.668 & $4.74\times 10^4$\\
20 & 0.806 & 0.836 & $4.29\times 10^5$ \\
10 & 0.748 & 0.844 & $8.99\times 10^5$ \\
5 & 0.614 & 0.821 & $1.63\times 10^6$\\
2.5 & 0.598 & 0.899 & $2.99\times 10^6$\\
1.25 & 0.470 & 0.900 & $4.24\times 10^6$
\end{tabular}
\end{table}

\section{Conclusion}
\label{sec:conclusion}

We have performed a thorough investigation of high-reflectivity and high-$Q$ resonances in thin subwavelength gratings illuminated by Gaussian beams at normal incidence. Finite grating size and beam collimation effects, which are known to limit the achievable peak reflectivity and $Q$-factor, are taken into account using FEM simulations and an optimization algorithm allows for efficiently extracting the optimal grating parameters under experimentally motivated constraints. The results of this investigation show that it should in principle be possible, by using optimized dual-period gratings, to improve by one to two orders of magnitude the linewidth-transmission loss product, which is the figure of merit relevant for resonant mirror-based microcavities currently investigated for various optomechanics and sensing applications~\cite{Naesby2018,Manjeshwar2023,Mitra2024}. We also note that the optimization approach used here may readily be applied to other photonic crystal structures whose performances cannot be captured by an infinite structure/plane wave approach and whose optimization requires computationally expensive simulations.


%
%
%
%

\section*{Data availability statement}

Data underlying the results presented in this article are not publicly available at this time but may be
obtained from the authors upon reasonable request.

\ack The authors acknowledge financial support from Novo Nordisk fonden.

\end{document}